\UseRawInputEncoding
\documentclass[journal=jacsat,manuscript=article]{achemso}
\usepackage{achemso}
\usepackage[version=3]{mhchem}
\usepackage[english]{babel}
\usepackage{filecontents}
\usepackage{graphicx}
\graphicspath{{./Figures/}}
\usepackage{courier}
\usepackage{hyperref,url}
\usepackage{amssymb,amsmath,amsthm,mathrsfs,comment,afterpage,accents}
\usepackage[capitalise]{cleveref}
\hypersetup{breaklinks,colorlinks=true,linkcolor=blue,citecolor=blue,urlcolor=blue,filecolor=blue}
\usepackage{mathtools}
\usepackage{braket}
\usepackage{mathrsfs}
\usepackage{courier}
\usepackage{multirow}
\usepackage{tabu}
\usepackage{color}
\usepackage{caption}
\usepackage{subcaption}
\usepackage{physics}
\usepackage{tablefootnote}

\usepackage[table,xcdraw]{xcolor}
\usepackage[numbers]{natbib}
\usepackage{xr} 
\usepackage{soul}
\usepackage[normalem]{ulem} 

\newcommand{\remove}[1]{}  

\makeatletter
\def\@dotsep{4.5}
\makeatother
\usepackage{dcolumn}
\usepackage{bm}
\usepackage{graphicx}

\renewcommand\vec\mathbf
\setkeys{acs}{maxauthors = 0}

\title{Assessment of the Performance of Density Functionals for Predicting Potential Energy Curves in Hydrogen Storage Applications}

\author{Srimukh Prasad Veccham}
\author{Martin Head-Gordon}
\email{mhg@cchem.berkeley.edu}
\affiliation{
Department of Chemistry, University of California, Berkeley, California 94720, USA
Chemical Sciences Division, Lawrence Berkeley National Laboratory, Berkeley, California 94720, USA
}

\begin{tocentry}
\includegraphics[width=8.2cm]{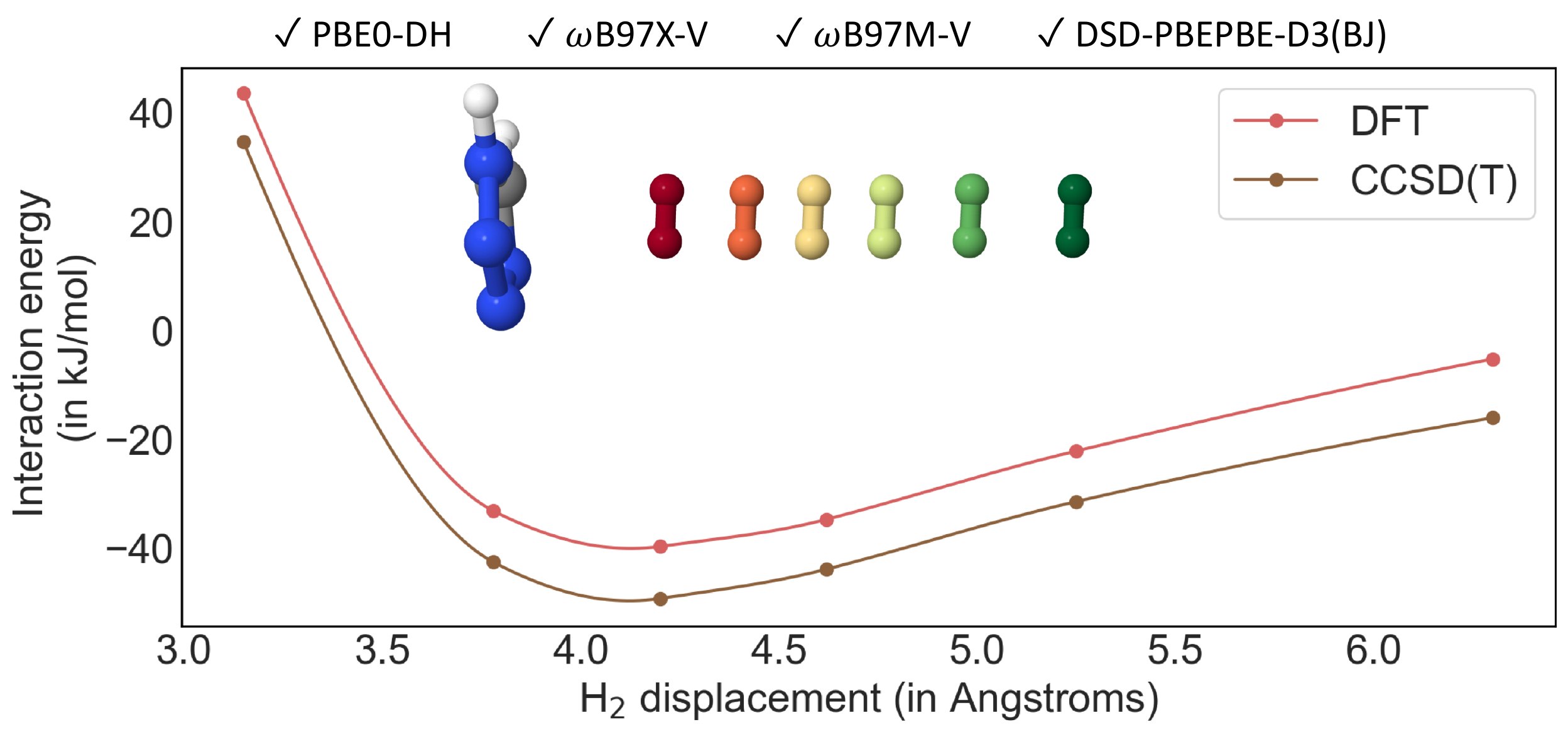}
\end{tocentry}

\begin{document}
\maketitle
\begin{abstract}
The availability of accurate computational tools for modeling and simulation is vital to accelerate  the discovery of materials capable of storing hydrogen (\ce{H2}) under given parameters of pressure swing and temperature.
Previously, we compiled the H2Bind275 dataset consisting of equilibrium geometries and assessed the performance of 55 density functionals over this dataset (Veccham, S. P.; Head-Gordon, M. \textit{J. Chem. Theory Comput.}, \textbf{2020}, \textit{16}, 4963--4982).
As it is crucial for computational tools to accurately model the entire potential energy curve (PEC), in addition to the equilibrium geometry, we have extended this dataset with 389 new data points to include two compressed and three elongated geometries along 78 PECs for \ce{H2} binding, forming the H2Bind78$\times$7 dataset.
Assessing the performance of 55 density functionals on this significantly larger and more comprehensive H2Bind78$\times$7 dataset, we have identified the best performing density functionals for \ce{H2} binding applications: PBE0-DH, $\omega$B97X-V, $\omega$B97M-V, and DSD-PBEPBE-D3(BJ).
Addition of Hartree Fock exchange improves the performance of density functionals, albeit not uniformly throughout the PEC.
We recommend the usage of $\omega$B97X-V and $\omega$B97M-V density functionals as they give good performance for both geometries and energies
In addition, we have also identified B97M-V and B97M-rV as the best semi-local density functionals for predicting \ce{H2} binding energy at its equilibrium geometry.
\end{abstract}

\section{Introduction}
Hydrogen (\ce{H2}) is a favorable substitute for fossil fuels as the only by-product of hydrogen fuel cell engines is water and the efficiency of a fuel cell is significantly higher than an internal combustion engine.
However, \ce{H2} is a light gas with low volumetric and gravimetric energy densities.
This poses a significant hurdle to storage and transportation of \ce{H2}.
Storing \ce{H2} reversibly in adsorbed form on porous materials is a promising solution to this problem.\cite{Takagi2004adsorptive, Thomas2007hydrogen,Murray2009hydrogen}
Ideally, such materials should adsorb \ce{H2} at high pressure and release it at low pressure so that the released \ce{H2} can be used for operating a fuel cell.
Designing materials with this property, while simultaneously not compromising on high volumetric and gravimetric storage capacities, is an active area of research.\cite{Park2012, Allendorf2018}

While multiple porous materials like Metal-Organic Frameworks (MOFs), Covalent Organic Frameworks (COFs), graphene, and other amorphous materials have been shown to adsorb \ce{H2}, none of these materials meet all the target criteria proposed by the U.S. Department of Energy for an ideal storage material.\cite{Allendorf2018}
As experimental synthesis and characterization of potential \ce{H2} storage materials is expensive and time-consuming, computational modeling and screening of materials has emerged as a viable alternative to it.\cite{Colon2014high, Thornton2017materials, Ahmed2019exceptional}
Computational techniques can be used in two different, potentially complementary ways.
First, molecular modeling can be used to understand the mechanism of \ce{H2} binding in different porous materials and this understanding can be used to systematically tune materials to achieve target properties.\cite{Kapelewski2014,Tsivion2017, Allendorf2018}
Second, computational techniques can be used to screen materials in a high-throughput manner to select only a handful of potentially viable materials for synthesis and characterization.

The ability of a material to store \ce{H2} is characterized by its usable capacity, which is defined as the amount of \ce{H2} stored at the high operating pressure that is released when the pressure is reduced to the low operating pressure.
Optimizing the usable capacity for typical fixed operating pressures of 5 bar and 100 bar gives an optimal value for the Gibbs free energy of adsorption ($\Delta G_{\text{ads}}$).
Assuming a correlation between enthalpy and entropy of adsorption in porous materials gives a range of $-15$ to $-25$ kJ/mol for the optimal value for enthalpy of adsorption ($\Delta H_{\text{ads}}$).\cite{Garrone2008, Bhatia2006, Bae2010}
The internal energy of binding, which is the largest component of $\Delta H_{\text{ads}}$,  can be computed using different quantum chemistry methods, including, but not limited to, density functional theory (DFT),\cite{Mueller2005density, Kapelewski2014, Tsivion2014, Tsivion2017} M\o ller-Plesset perturbation theory (MP2),\cite{Cabria2008hydrogen,Cabria2011simulation, Niaz2014theoretical} and different variants of coupled-cluster theory.\cite{Kocman2015choosing, Ma2015computational, Veccham2020}
Each of these methods have different accuracies and computational costs associated with them.

DFT, scaling as $\mathcal{O}(N^3)$ ($N$ is the number of basis functions in the system), can provide a reasonable balance between cost and accuracy of computing \ce{H2} binding energy.
However, as the exact density functional remains unknown, different density functional approximations (DFAs), proposed in lieu of the exact density functional, provide varying accuracies for different chemical systems and/or properties computed.\cite{Mardirossian2016}
In order to address this problem, we adopted a two-pronged approach.\cite{Veccham2020}
(1) We compiled the H2Bind275 dataset that consists of \ce{H2}(s) interacting with binding motifs representative of different porous materials known for \ce{H2} adsorption. This dataset consists only of equilibrium geometries, that is, \ce{H2}(s) are located at the minimum of the potential energy curve (PEC) with respect to the binding site. We computed highly accurate reference interaction energies using coupled-cluster singles, doubles, and perturbative triples (CCSD(T)) extrapolated to the complete basis set limit for this dataset.
(2) We assessed the performance of 55 DFAs and identified the best performing density functionals for this dataset. In addition, we also identified inexpensive semi-local density functionals which give very good performance for low computational cost and are suitable for in silico high-throughput screening purposes.

The H2Bind275 dataset, consisting of 275 data points, provides a balanced representation of different \ce{H2} binding mechanisms like polarization, charge transfer interaction, and dispersion.\cite{Sillar2009, Tsivion2014, Koizumi2019hydrogen}
It also captures the chemical diversity of binding motifs that \ce{H2} interacts with in porous frameworks.
This dataset assesses the ability of density functionals to reproduce \ce{H2} binding energies at the minima of the PEC.
However, as DFAs are routinely used for geometry optimizations and molecular dynamics simulations either directly or indirectly (by generating reference data for training force fields), they should also be able to reproduce the entire PEC which would ensure accurate nuclear gradients as required for geometry optimization and molecular dynamics simulations.
A strategy of assessing the performance of DFAs for PECs has been previously employed for other non-covalent interaction energy datasets like S22, S66, and A24.
The S22x5 dataset\cite{Grafova2010comparative} was created from the S22 dataset\cite{Jurevcka2006benchmark} by including geometries that are shortened and elongated along a well-defined interaction coordinate.
Similarly, the S66x8\cite{Rezac2011} and A21x12\cite{Witte2015} extended datasets were created from the S66 and A24 datasets.\cite{Rezac2013describing}

In order to address this issue for \ce{H2} storage, we have extended the H2Bind275 dataset to include geometries that are located at five different points on 78 separate PECs, not just the minimum.
This extended dataset, hereafter referred to as the H2Bind78$\times$7 dataset, was generated by shortening and stretching the distance between \ce{H2} and the binding motif.
The reference interaction energies were computed using CCSD(T) extrapolated to the complete basis set (CBS) limit using the same strategy outlined in Ref.~\citenum{Veccham2020}.
The performance of 55 DFAs were assessed using regularized relative errors metrics by appropriately weighing the error coming from different points on the PECs.
We have analyzed the performance of these DFAs for the extended dataset by comparing and contrasting it with the performance of the original equilibrium H2Bind275 dataset.

This paper is organized as follows. The H2Bind78$\times$7 dataset is introduced and the protocol for computing reference interaction energies is discussed.
All the density functionals chosen for assessment in this work are briefly introduced and classified.
The distribution of the reference interaction energies at different points on the PEC is outlined.
The performance of DFAs for predicting \ce{H2} interaction energies across the PEC is discussed and contrasted with their performance for the previous H2Bind275 dataset.
The performance of DFAs for predicting equilibrium geometries and interaction energies at equilibrium geometries is explored.
The best DFAs for predicting \ce{H2} binding energies are recommended while considering their computational cost.

\section{Computational details}
\subsection{H2Bind78$\times$7 dataset}

\begin{table}[]
\caption{Number of geometries and data points by chemical categories for the H2Bind78$\times$7 dataset}
\label{tab:H2Bind162x6_dataset}
\resizebox{\textwidth}{!}{%
\begin{tabular}{cccccc}
\hline
                     & s-block ions & salts & organic ligands & transition metals & total \\ \hline
geometries           & 19           & 13    & 5               & 41                & 78    \\
data points at PEC minimum     & 38           & 26    & 10              & 82                & 156   \\
data points not at PEC minimum & 95           & 65    & 25              & 204\tablefootnote{One data point excluded due to convergence issues}               & 389   \\
H2Bind78$\times$7            & 133          & 91    & 35              & 286               & 545   \\ \hline
\end{tabular}%
}
\end{table}

The H2Bind275 dataset consists of 275 \ce{H2} interaction energies but only 86 unique geometries as many of them have multiple \ce{H2}s.
For example, the geometry of \ce{CaCl2-(H2)4} has four hydrogen molecules bound to \ce{CaCl2} contributing four data points to the H2Bind275 dataset.
The H2Bind78$\times$7 dataset was generated by starting from a subset of the original H2Bind275 dataset.
This subset was created by choosing only 78 unique geometries and computing their interaction energies adiabatically using the method outlined in Ref.~\citenum{Veccham2020}.
All of these geometries are located on the PEC at their respective minima.
The adiabatic interaction energy, which relaxes the geometries of the binding motif and \ce{H2}, was chosen as it is closest to experimentally measurable values.
For each minimum geometry, five additional geometries were generated by compressing and elongating the distance between the binding motif and the center of mass of \ce{H2} (denoted by $r_{\text{eq}}$).
For geometries containing multiple \ce{H2}s bound to a single binding moiety, compressed and elongated geometries were generated for only one of the \ce{H2}s.
This step was necessary in order to maintain low redundancy in the dataset and make its size manageable.
In addition to this, the interaction energy at the minimum of the PEC was also computed using vertical interaction energy method.
In total, this dataset contains 78 adiabatic and 78 vertical interaction energies (a total of 156 data points) located at the PEC minimum.

In this work, two compressed geometries ($0.75r_{\text{eq}}$ and $0.9r_{\text{eq}}$) and three elongated geometries ($1.1r_{\text{eq}}$, $1.25r_{\text{eq}}$, and $1.5r_{\text{eq}}$) were considered.
These distances were chosen as they are representative of the PEC in both the compressed and elongated regimes.
In a porous material, \ce{H2} interacts with not only its primary binding site but also has secondary interactions with other components of the framework.
The binding distances of \ce{H2} to its secondary interaction sites of the porous material are often longer than their corresponding equilibrium distances.
As a consequence of this, when modeling \ce{H2} in a porous material, the elongated portion of the PEC is sampled more than the compressed part.
Additionally, the compressed portion of the PEC is usually significantly higher in energy (repulsive if compressed enough), and is sampled less often in a molecular dynamics or Monte Carlo simulation.
Hence, DFAs should be able to reproduce the elongated portion of the PEC more faithfully than the compressed portion.
We have included more data points in the elongated regime than the compressed regime in order to underscore its relative importance.
As shown in Table~\ref{tab:H2Bind162x6_dataset}, the number of non-equilibrium data points is roughly $2.5$ times the number of data points at equilibrium.
In total, counting both the equilibrium and non-equilibrium data points, this H2Bind78$\times$7 dataset consists of $545$ \ce{H2} interaction energies with representative binding motifs.

\begin{table}[]
    \caption{All 78 geometries in the H2Bind78$\times$7 dataset categorized by chemical identity of the binding motif.}
    \label{tab:dataset_enumerated}
    \resizebox{\textwidth}{!}{%
    \begin{tabular}{c|c|c|c}
    \hline
        s-block ions & salts & organic ligands & transition metals \\ \hline
        \ce{Li+-(H2)_n},  &  \ce{AlF3-H2} & benzene\ce{-H2} & \ce{MX-H2}, X$=$H, F, Cl; M$=$Cu, Ag, Au  \\
         $n=1,2,3,4,5,6$  &  \ce{CaF2-(H2)_n},  & phenol\ce{-H2} & \ce{CoF3-H2} \\
        \ce{Na+-(H2)_n},  &  $n=1,2,3,4$  & pyrrole\ce{-H2} & \ce{Cu(OMe)-H2} \\
        $n=1,2,3$ &  \ce{CaCl2-(H2)_n},  &  butene\ce{-H2} & \ce{CuCN-H2} \\
        \ce{Mg^{2+}-(H2)_n},   & $n=1,2,3,4$   & tetrazole\ce{-H2} & \ce{Sc+-(H2)_n}, \ce{V+-(H2)_n}, $n=3,4$ \\ 
        $n=1,2,3,4$   & \ce{MgF2-(H2)_n},   &  & \ce{Ti+-(H2)_n}, $n=2,4$\\ 
        \ce{Ca^{2+}-(H2)_n}, &  $n=1,2,3,4$ &  & \ce{Cr+-(H2)_n}, \ce{Mn+-(H2)_n}, $n=1,2,3,4$ \\ 
         $n=1,2,3,4,5,6$   &   &  & \ce{Fe+-(H2)_n}, $n=1,2,3,4$ \\ 
          &   &  &  \ce{Co+-(H2)_n}, \ce{Ni+-(H2)_n}, $n=1,2$ \\ 
          &   &  &  \ce{Cu+-(H2)_n}, $n=1,2,3$ \\ 
          &   &  &  \ce{Zn+-(H2)_n}, $n=1,2,3,4$ \\ \hline
    \end{tabular} }
\end{table}
This dataset, like the H2Bind275 dataset, can also be divided into categories based on the chemical nature of the binding motif as shown in Table~\ref{tab:dataset_enumerated}: (1) s-block ions: consisting of group 1 and group 2 bare metal cations with unscreened charge binding one or multiple \ce{H2}s, (2) salts: consisting of small inorganic salts like \ce{AlF3}, \ce{CaCl2}, and \ce{MgF2} binding one or multiple \ce{H2}s, (3) organic ligands: comprising of small aliphatic and aromatic molecules binding one \ce{H2}, (4) transition metals: including small transition metal complexes and 3d transition metal cations binding one or multiple \ce{H2}s.
Each of these categories is also representative of various mechanisms of \ce{H2} binding found in porous materials.
For example, \ce{H2} in the organic ligands category is mostly dispersion-bound.
The s-block metals category binds \ce{H2} using a combination of electrostatic and forward charge transfer (\ce{H2} $\rightarrow$ metal) interactions.\cite{Tsivion2014}
This dataset captures both chemical and mechanistic diversity encountered in \ce{H2} binding to porous materials.
For a detailed discussion about different chemical categories in this dataset, we refer readers to Ref.~\citenum{Veccham2020}.

\subsection{Reference Binding Energies}
Calculation of accurate reference interaction energies is an important task in compiling a dataset.
Reference interaction energies were computed using coupled-cluster theory with singles, doubles, and perturbative triples (CCSD(T))\cite{Raghavachari1989} extrapolated to the complete basis set limit.
Inspired by the success of composite extrapolation methods\cite{Tajti2004, DeYonker2006, Karton2006, Curtiss2007} for computing highly accurate reference values, we have developed our own composite extrapolation method using focal point analysis\cite{East1993,Csaszar1998} for computing accurate reference \ce{H2} binding energies:
\begin{align}
    E_{\text{ref}} &= E_{\text{HF/5Z}} + E_{\text{MP2/QZ}\rightarrow \text{5Z}} + \delta E_{\text{CCSD(T)/TZ}} + \delta E^{\text{core}}_{\text{MP2/TZ}}  \label{eq::composite_CC} \\
    \delta E_{\text{CCSD(T)/TZ}} &= E_{\text{CCSD(T)/TZ}} - E_{\text{MP2/TZ}} \label{eq::composite_fpa} \\
    \delta E^{\text{core}}_{\text{MP2/TZ}} &= E^{\text{core=0}}_{\text{MP2/TZ}} - E^{\text{core=n}}_{\text{MP2/TZ}} \label{eq::composite_fc}
\end{align}
Here, $E_{\text{ref}}$ is the reference energy computed using the composite method, $E_{\text{HF/5Z}}$ is the Hartree Fock energy computed using a basis set of quintuple-zeta (5Z) quality, $E_{\text{MP2/QZ}\rightarrow\text{5Z}}$ is the MP2 correlation energy extrapolated to the complete basis set limit with the 2-point extrapolation formula\cite{Helgaker1997} using correlation energies computed with quadruple-zeta (QZ) and quintuple-zeta quality basis sets, and $\delta E_{\text{CCSD(T)/TZ}}$ is the difference between the CCSD(T) and MP2 correlation energies computed with a triple-zeta quality basis set.
$\delta E^{\text{core}}_{\text{MP2/TZ}}$ is the core-valence contribution to the correlation energy computed as the difference between MP2 correlation energies with ($E^{\text{core=n}}_{\text{MP2/TZ}}$) and without ($E^{\text{core=0}}_{\text{MP2/TZ}}$) the frozen-core approximation.
This composite method for computing reference \ce{H2} interaction energies ensures that the effect of higher-order excitations neglected in CCSD(T) are sufficiently small.
It also ensures that the basis set incompleteness errors are small and that both HF and extrapolated correlation energy components are of complete basis set limit quality.
Further details of this scheme can be found in Ref.~\citenum{Veccham2020}.

The cc-pVnZ\cite{Dunning1989,Woon1993} ($n=$T, Q, or 5) family of basis sets was used for all the HF and correlation energy calculations when core electrons were not included in the correlation calculations.
cc-pCVnZ\cite{Woon1995,Peterson2002} ($n=$T, Q, or 5) family of basis sets were employed when some or all of the core electrons were included in the correlation calculations.
For transition metals, the cc-pwCVnZ\cite{Balabanov2005} ($n=$T, Q, or 5) series of basis sets was used with a neon core excluded in all correlation energy computations.

\subsection{Density Functional Approximations}
55 DFAs, including all the commonly used density functionals, were chosen to perform a thorough assessment.
We have also included DFAs that have previously shown very good performance for a range of non-covalent interaction energy prediction problems represented by multiple datasets.\cite{Mardirossian2017}
Based on the different quantities DFAs depend on, they are categorized into rungs of the metaphorical Jacob's ladder.\cite{Perdew2005}
From the first rung of the Jacob's ladder, in which DFAs depend only on electron density, SVWN5\cite{Dirac1930note, Vosko1980} and SPW92\cite{Dirac1930note, Perdew1992} DFAs were chosen.
From the second rung called Generalized Gradient Approximation (GGA), 12 different DFAs were chosen: the PBE family and its variants (PBE,\cite{Perdew1996} PBE-D3(0),\cite{Grimme2010} RPBE,\cite{Hammer1999} revPBE,\cite{Zhang1998} and revPBE-D3(op)\cite{Witte2017}), BLYP\cite{Becke1988, Lee1988} and BLYP-D3(op)\cite{Witte2017}, dispersion-corrected variants of B97\cite{Becke1997} (BLYP-D3(0)\cite{Grimme2010} and BLYP-D3(BJ)\cite{Grimme2011}), PW91,\cite{Perdew1992a} and GAM.\cite{Yu2015}
From the meta-GGA rung, the different variants of TPSS (TPSS,\cite{Tao2003} TPSS-D3(BJ),\cite{Grimme2011} and revTPSS\cite{Perdew2009}), SCAN\cite{Sun2015strongly} and its dispersion-corrected version SCAN-D3(BJ)\cite{Brandenburg2016benchmark}, MS2\cite{Sun2013} and MS2-D3(op),\cite{Witte2017} the combinatorially-optimized B97M-V\cite{Mardirossian2015} and B97M-rV\cite{Sabatini2013,Mardirossian2016a} were chosen.
In addition, mBEEF\cite{Wellendorff2014} and the semi-local Minnesota functionals M06-L\cite{Zhao2006} and MN15-L\cite{Yu2016a} were also included in the assessment. 
Rung four DFAs, containing HF exchange, are generally more accurate than semi-local functionals as they partially alleviate the problem of self-interaction error.
In this work, global hybrid density functionals like B3LYP\cite{Becke1993c} and B3LYP-D3(0),\cite{Grimme2011} PBE0\cite{Adamo1999} and PBE0-D3(BJ),\cite{Grimme2011} TPSSh\cite{Staroverov2003} and TPSSh-D3(BJ),\cite{Grimme2011} the M06 family of density functionals (M06,\cite{Zhao2008} M06-2X,\cite{Zhao2008} M06-2X-D3(0),\cite{Grimme2010} and revM06\cite{Wang2018}), MVSh,\cite{Sun2015semilocal} and SCAN0\cite{Hui2016} which is the hybrid variant of SCAN are included.
Range-separated hybrids, which are hybrid functionals containing DFT exchange and some HF exchange in the short-range and only HF exchange in the long range, included in this study are $\omega$B97X-D,\cite{Chai2008} $\omega$B97X-D3,\cite{Lin2012} $\omega$B97X-V,\cite{Mardirossian2014a}  $\omega$B97M-V,\cite{Mardirossian2016} M11\cite{Peverati2011} and its revised version revM11.\cite{Verma2019}
Two screened exchange density functionals (HSE-HJS\cite{Krukau2006,Henderson2008} and MN12-SX\cite{Peverati2012}), which contain DFT exchange in the short range and attenuated HF exchange in the long range are also included.
Double hybrid density functionals, which are at the top the Jacob's ladder classification, contain some percentage of correlation energy from wavefunction methods.
These DFAs are characterized by their superior accuracy and increased computational cost in comparison to semi-local and hybrid DFAs.
We have included seven double hybrid density functionals in this study: B2PLYP-D3(BJ),\cite{Grimme2006} XYG3,\cite{Zhang2009} XYGJ-OS,\cite{Zhang2011} PBE0-DH,\cite{Bremond2011}, PTPSS-D3(0),\cite{Goerigk2011} DSD-PBEPBE-D3(BJ),\cite{Kozuch2013} and $\omega$B97M(2).\cite{Mardirossian2018a}  

The def2-QZVPPD\cite{Rappoport2010a} basis set was used for all DFA calculations with a quadrature grid of 99 Euler-MacLaurin radial points and 590 Lebedev angular points for integrating the exchange-correlation contribution.
SG-1\cite{Gill1993} integration grid was used for integrating the VV10 component.
The choice of core for frozen core approximation and employment of density fitting approximation for computing MP2 correlation energy in double hybrid density functionals is discussed in Table~S1.
All the PECs were interpolated using the one-dimensional Akima interpolator.\cite{Akima1970new}
All computations were performed using Q-Chem 5.\cite{Shao2015}

\section{Results and Discussion}
\subsection{H2Bind78$\times$7 dataset}
\begin{figure}
    \centering
    \includegraphics[width=\linewidth]{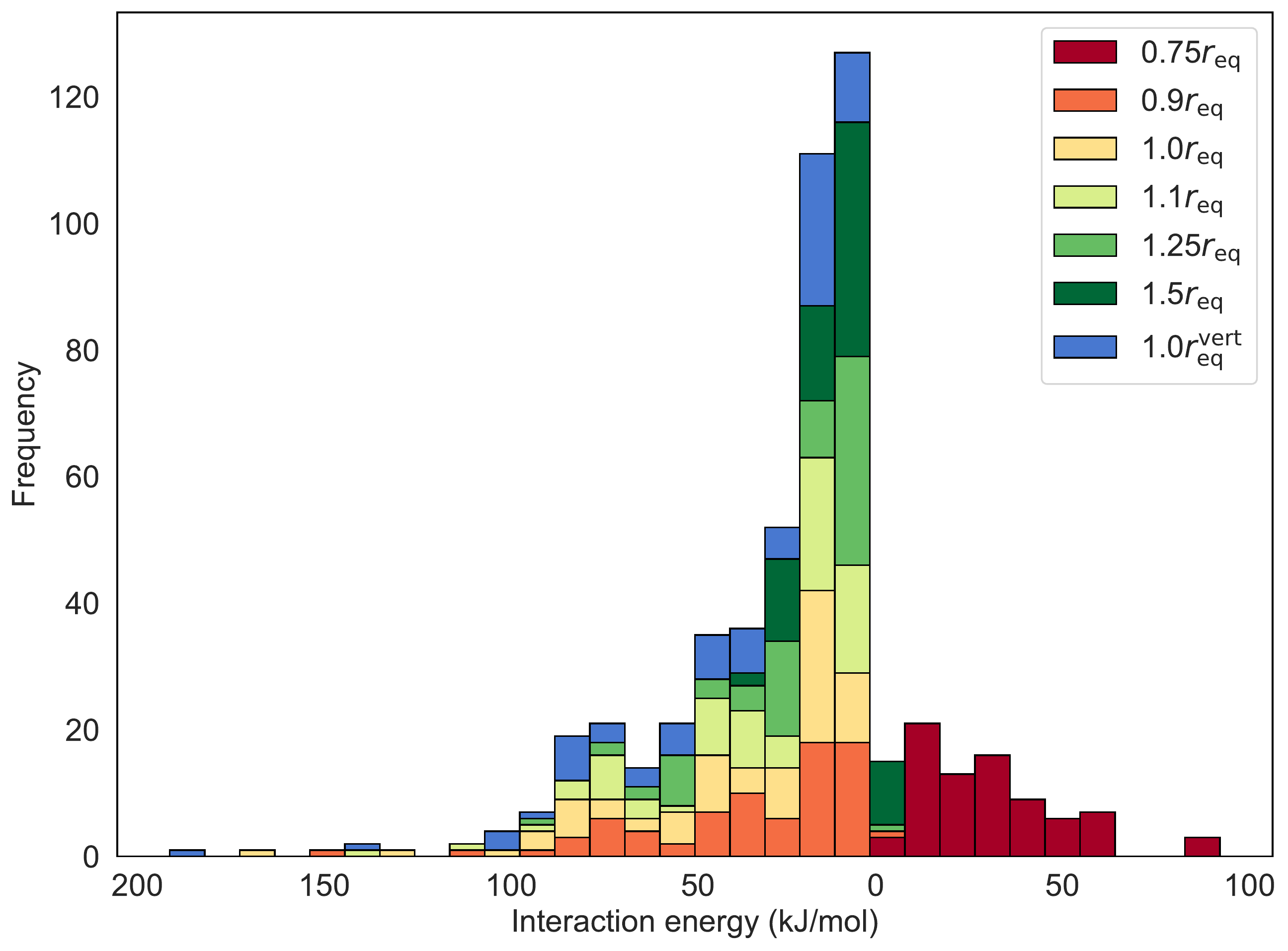}
    \caption{Distribution of coupled-cluster reference interaction energies separated by location on the potential energy curve. The reference vertical interaction energy at equilibrium ($1.0r_{\text{eq}}^{\text{vert}}$) is also shown.} 
    \label{fig:benchmark_hist}
\end{figure}
Typically, the coupled cluster reference \ce{H2} interaction energy with the binding motif is strongest at equilibrium, that is at $r_{\text{eq}}$.
This implies that the geometries optimized using $\omega$B97M-V/def2-TZVPD\cite{Rappoport2010a} are also close to the CCSD(T)/CBS minima. 
Fig.~\ref{fig:benchmark_hist} shows the distribution of interaction energies for the entire H2Bind78$\times$7 dataset consisting of 545 data points.
The equilibrium and geometries near equilibrium ($0.9r_{\text{eq}}, 1.0r_{\text{eq}}$, $1.0r_{\text{eq}}^{\text{vert}}$, and $1.1r_{\text{eq}}$) have attractive interaction energies, with most of them smaller than 100 kJ/mol in magnitude.
Geometries that are stretched by 25\% ($1.25r_{\text{eq}}$) are still attractive in nature, but most interaction energies are smaller than 60 kJ/mol in magnitude.
Geometries stretched by 50\% of their equilibrium distance are bound only weakly with a median binding energy of $-6.9$ kJ/mol.
At the other extreme of the PEC, geometries that are compressed by 25\% ($0.75r_{\text{eq}}$) are mostly repulsive with a median interaction energy of $+29.7$ kJ/mol.
This geometry was also chosen in order to sample the repulsive part of the PEC and assess how accurately different density functionals can reproduce it.\cite{Taylor2016blind}

The range of interaction energies covered by each PEC is also very large.
The coinage metal containing species are the strongest binders, as illustrated by the extreme example of \ce{AuF} which binds \ce{H2} with an interaction energy of $-161.8$ kJ/mol at equilibrium and interacts with \ce{H2} with an energy of $+87.6$ kJ/mol (repulsive) at $0.75r_{\text{eq}}$, thus spanning an interaction range of $249.4$ kJ/mol.
Data points in the organic category have the smallest ranges (average range is $19$ kJ/mol).
A typical PEC of \ce{H2} interacting with a binding moiety has the shape of a Morse potential.
However, there is considerable variation in the well depth, well width, and decay in the long range for different chemical species.
This variation can provide some clues into the dominant mechanism of interaction.
For example, the \ce{AuCl} binding motif interacts with one \ce{H2} with an interaction energy of $-123.7$ kJ/mol at equilibrium which decays to $-21.3$ kJ/mol at $1.5 r_{\text{eq}}$ ($82.7\%$ decrease).
This is a sharp decay in the interaction energy in comparison to the \ce{Mg^2+} case.
In the \ce{Mg^2+} interacting with one \ce{H2} case, the interaction energy at equilibrium is $-97.8$ kJ/mol in comparison to $-37.7$ kJ/mol at $1.5 r_{\text{eq}}$ ($61.4\%$ decrease).
This suggested that the dominant mechanism of interaction in \ce{Mg^2+} case is longer-ranged (like permanent electrostatics) than \ce{AuCl} which is dominated by orbital controlled short-ranged interactions like charge transfer.

\subsection{Performance of Density Functional Approximations on PECs}

\begin{table}[]
\caption{Regularized mean absolute percentage error (RegMAPE) and root mean squared error (RMSE; in kJ/mol) of all DFAs considered in this work for the entire H2Bind78$\times$7 dataset.}
\label{tab:err_regmape_rmse}
\resizebox{0.78\textwidth}{!}{%
\begin{tabular}{|c|cc|cc|}
\hline
Rank	&	DFA	&	RMSE (kJ/mol)	&	DFA	&	RegMAPE (\%)	\\ \hline
1	&	PBE0-DH	&	2.9	&	PBE0-DH	&	5.0	\\
2	&	DSD-PBEPBE-D3(BJ)	&	3.7	&	$\omega$B97X-V	&	5.4	\\
3	&	$\omega$B97X-V	&	4.0	&	$\omega$B97M-V	&	6.3	\\
4	&	$\omega$B97X-D	&	4.1	&	DSD-PBEPBE-D3(BJ)	&	6.3	\\
5	&	PBE0	&	4.2	&	XYGJ-OS	&	6.9	\\
6	&	MVSh	&	4.3	&	$\omega$B97M(2)	&	7.4	\\
7	&	HSE-HJS	&	4.3	&	PBE0	&	7.6	\\
8	&	$\omega$B97M-V	&	4.5	&	HSE-HJS	&	7.6	\\
9	&	XYGJ-OS	&	4.6	&	B2PLYP-D3(BJ)	&	8.2	\\
10	&	$\omega$B97X-D3	&	4.8	&	XYG3	&	8.5	\\
11	&	XYG3	&	4.8	&	$\omega$B97X-D	&	9.0	\\
12	&	PTPSS-D3(0)	&	4.9	&	B97M-rV	&	9.0	\\
13	&	$\omega$B97M(2)	&	5.1	&	B97M-V	&	9.1	\\
14	&	PBE0-D3(BJ)	&	5.2	&	SCAN0	&	9.1	\\
15	&	MN15	&	5.7	&	PTPSS-D3(0)	&	9.3	\\
16	&	B2PLYP-D3(BJ)	&	5.8	&	$\omega$B97X-D3	&	9.8	\\
17	&	SCAN0	&	5.8	&	MVSh	&	10.2	\\
18	&	TPSSh	&	6.0	&	TPSSh	&	11.3	\\
19	&	revM11	&	6.3	&	PBE0-D3(BJ)	&	11.5	\\
20	&	mBEEF	&	6.7	&	M11	&	12.0	\\
21	&	revM06	&	6.9	&	revTPSS	&	12.0	\\
22	&	B3LYP	&	7.4	&	revM06	&	12.4	\\
23	&	revTPSS	&	7.5	&	TPSS	&	13.6	\\
24	&	B3LYP-D3(0)	&	7.5	&	TPSSh-D3(BJ)	&	13.8	\\
25	&	B97M-V	&	7.6	&	B3LYP-D3(0)	&	14.0	\\
26	&	B97M-rV	&	7.6	&	oTPSS-D3(BJ)	&	14.6	\\
27	&	TPSS	&	7.7	&	TPSS-D3(BJ)	&	14.7	\\
28	&	oTPSS-D3(BJ)	&	7.8	&	MN15	&	15.3	\\
29	&	TPSSh-D3(BJ)	&	8.0	&	PBE	&	15.3	\\
30	&	MN15-L	&	8.0	&	revM11	&	15.3	\\
31	&	revPBE-D3(op)	&	8.5	&	BLYP-D3(op)	&	15.4	\\
32	&	TPSS-D3(BJ)	&	8.5	&	B3LYP	&	15.5	\\
33	&	revPBE	&	9.1	&	SCAN	&	15.6	\\
34	&	MN12-SX	&	9.2	&	MN12-SX	&	16.2	\\
35	&	RPBE	&	9.3	&	SCAN-D3(BJ)	&	16.3	\\
36	&	M11	&	9.4	&	M06	&	16.9	\\
37	&	B97-D3(BJ)	&	9.7	&	revPBE-D3(op)	&	17.0	\\
38	&	BLYP-D3(op)	&	9.8	&	PW91	&	17.4	\\
39	&	M06	&	9.9	&	MS2	&	17.4	\\
40	&	BLYP	&	10.1	&	mBEEF	&	17.4	\\
41	&	PBE	&	10.3	&	BP86-D3(BJ)	&	18.4	\\
42	&	M06-L	&	10.5	&	M06-2X	&	19.3	\\
43	&	BP86-D3(BJ)	&	10.9	&	M06-2X-D3(0)	&	19.8	\\
44	&	B97-D3(0)	&	11.0	&	MN15-L	&	19.8	\\
45	&	PBE-D3(0)	&	11.0	&	PBE-D3(0)	&	20.1	\\
46	&	PW91	&	11.2	&	MS2-D3(op)	&	20.3	\\
47	&	GAM	&	11.8	&	M06-L	&	20.7	\\
48	&	MS2	&	12.1	&	RPBE	&	20.9	\\
49	&	MS2-D3(op)	&	12.4	&	BLYP	&	22.1	\\
50	&	SCAN	&	12.9	&	revPBE	&	23.4	\\
51	&	SCAN-D3(BJ)	&	13.2	&	B97-D3(BJ)	&	24.2	\\
52	&	M06-2X	&	13.4	&	GAM	&	24.7	\\
53	&	M06-2X-D3(0)	&	13.4	&	B97-D3(0)	&	28.3	\\
54	&	SPW92	&	32.6	&	SPW92	&	63.0	\\
55	&	SVWN5	&	32.7	&	SVWN5	&	63.0	\\ \hline
\end{tabular}%
}
\end{table}
We will discuss the performance of DFAs using multiple error metrics.
Each of these metrics gives different weights to different aspects of the dataset.
First, we will discuss the performance of DFAs using the root mean square error (RMSE) metric which gives equal importance to all data points in the H2Bind78$\times$7 dataset.
The RMSE of all 55 DFAs assessed in this work is shown in Table~\ref{tab:err_regmape_rmse}.
The non-empirical double hybrid functional with just two fixed parameters, PBE0-DH, gives the least RMSE of 2.9 kJ/mol.
The second best DFA is another double hybrid DSD-PBEPBE-D3(BJ) with an RMSE of 3.7 kJ/mol.
In comparison to the earlier H2Bind275 dataset where DSD-PBEPBE-D3(BJ) was ranked fourth, it performs relatively better for the H2Bind78$\times$7 dataset moving up by two places.\cite{Veccham2020}
This is closely followed by $\omega$B97X-V and $\omega$B97X-D, both of which show a similar RMSEs of $4.0$ and $4.1$ kJ/mol respectively.
Another trend seen in the H2Bind275 dataset that is transferable to the H2Bind78$\times$7 dataset is that the best performing DFA in each rung of the Jacob's ladder performs better than the best performing functional in rung directly below it. 
The least RMSE DFA in each rung also remains the same: SPW92 for LDA, revPBE-D3(op) for GGAs, mBEEF for meta-GGAs, $\omega$B97X-V in the hybrids rung, and PBE0-DH in the double hybrids rung.
The ranking of $\omega$B97M-V deteriorates in the extended dataset in comparison to the previous H2Bind275 dataset.
Another interesting observation is the improvement in the ranking of the MN15 density functional which is ranked 15\textsuperscript{th} in the H2Bind78$\times$7 dataset with an RMSE of $5.7$ kJ/mol (MN15 was ranked 25\textsuperscript{th} with an RMSE of $6.3$ kJ/mol in the H2Bind275 dataset).
The relative performance of B97M-V and B97M-rV (ranked 25\textsuperscript{th} and 26\textsuperscript{th}) in the H2Bind78$\times$7 dataset remains comparable to their performance in the H2Bind275 dataset.
We also note that commonly used density functionals like M06-2X and M06-2X-D3(0) and the recently developed density functionals like SCAN and SCAN-D3(BJ) show very large RMSEs.

\begin{figure}
    \centering
    \includegraphics[width=\linewidth]{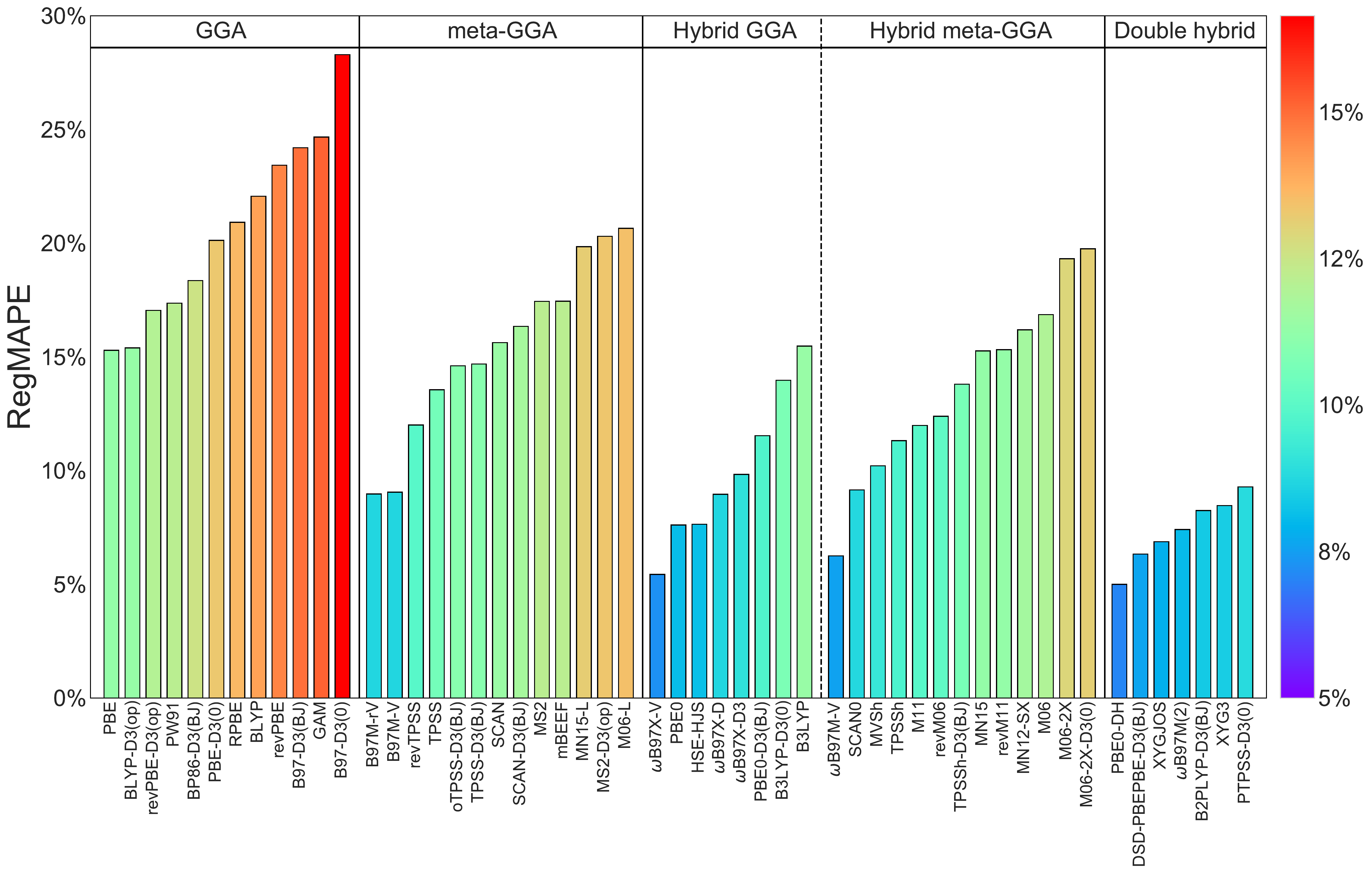}
    \caption{Performance of density functional approximations for the H2Bind78$\times$7 dataset assessed using regularized mean absolute percentage error (RegMAPE). The LDA density functionals, SPW92 and SVWN5, are not included in this figure and show a large RegMAPE of 63.0\%.}
    \label{fig:dft_performance_regmape}
\end{figure}
The reference interaction energies in the H2Bind78$\times$7 dataset span a very large range: from $-189.0$ to 92.1 kJ/mol.
However, for \ce{H2} storage applications between 5 and 100 bar, interaction energies in the $-15$ to $-25$ kJ/mol range would be ideal.\cite{Garrone2008, Bhatia2006, Bae2010}
A good error metric should give more weight to data points in this range by considering the following aspects:
\begin{enumerate}
    \item The H2Bind78$\times$7 dataset contains many model binding motifs, each of them binding \ce{H2} with different interaction energies at their corresponding equilibrium geometry. Binding motifs that bind \ce{H2} with an interaction energy in the range of $-15$ to $-25$ kJ/mol should have larger weights.
    \item Each binding motif contributes six data points: two compressed data points, three elongated data points, and one data point at equilibrium. Equilibrium geometries should be given larger weight than the non-equilibrium ones.
\end{enumerate}
The regularized mean absolute percentage error (RegMAPE) was formulated in Ref.~\citenum{Veccham2020} in order to satisfy requirement (1).
RegMAPE uses percentage error in the $-15$ to $-25$ kJ/mol range, regularized percentage error (in order to avoid small denominators) for interaction energies weaker than $-15$ kJ/mol, and absolute error for interaction energies stronger than $-25$ kJ/mol.
The error metrics in neighboring ranges are also smoothly interpolated.
For the same amount of error, as percentage error is much larger in magnitude than absolute error, the RegMAPE error metric is able to satisfy criterion (1).
For example, an error of 5 kJ/mol for a reference interaction energy of 100 kJ/mol will contribute 5 units to the total error while the same error for a reference interaction energy of 20 kJ/mol will contribute 25 units to the total error.
For a given binding motif, the equilibrium geometry should be given more weight as it represents the \ce{H2} interaction with the primary binding site: the main lever to tune while designing binding sites.
As non-equilibrium geometries are higher in energy, they would be encountered less frequently in a molecular dynamics or Monte Carlo simulation.
Hence, lower weight for non-equilibrium geometries is achieved by using the equilibrium regularization value for non-equilibrium geometries as well.
As the equilibrium geometry always has a stronger interaction energy, the regularized value of its reference interaction energy, $\tilde{E}(r_{\text{eq}})$, will be larger in magnitude in comparison to the regularized values of non-equilibrium interaction energies ($\tilde{E}(\alpha r_{\text{eq}}), \alpha \neq 1.0$).
The large magnitude of the denominator will give a smaller weight to the errors of non-equilibrium geometries in comparison to the equilibrium one.
RegMAPE for equilibrium and non-equilibrium geometries as defined in Eq.~\eqref{eq:regMUPE_noneq} satisfies requirement (2).
\begin{align}
    \Delta E( \alpha r_{\text{eq}}) &= \frac{ E^{\text{DFA}}( \alpha r_{\text{eq}})- E^{\text{ref}}( \alpha r_{\text{eq}})}{ \tilde{E}(r_{\text{eq}})}, \quad \alpha \in \{0.75, 0.9, 1.0, 1.1, 1.25, 1.5 \} \label{eq:regMUPE_noneq}
\end{align}
where $\Delta E( \alpha r_{\text{eq}})$ is the RegMAPE, $ E^{\text{DFA}}( \alpha r_{\text{eq}})$ and  $ E^{\text{ref}}( \alpha r_{\text{eq}})$ are the DFA and reference interaction energies at $\alpha r_{\text{eq}}$ geometry, and $ \tilde{E}(r_{\text{eq}})$ is the regularized interaction energy for the equilibrium geometry.
As the vertical interaction energy is located at the minimum of the PEC, the error in this data point is regularized using the vertical reference interaction energy.

The performance of DFAs assessed by the RegMAPE error metric is shown in Fig.~\ref{fig:dft_performance_regmape} and Table~\eqref{tab:err_regmape_rmse}.
While there are some similarities in the relative ordering of density functionals for RegMAPE and RMSE error metrics, there are also noteworthy differences.
Again, PBE0-DH shows the best performance with the least RegMAPE of 5.0\%.
It is followed by the $\omega$B97X-V and $\omega$B97M-V density functionals which have RegMAPEs of 5.4\% and 6.3\% respectively.
The DSD-PBEPBE-D3(BJ) density functional, which was the best performing density functional in the H2Bind275 dataset with RegMAPE of 4.9\%, is the fourth best performing DFA for the H2Bind78$\times$7 dataset.
The small decline in the performance of DSD-PBEPBE-D3(BJ) can be attributed to its poor performance for the non-equilibrium geometries as shown in Table~S2.
Another noteworthy decline in performance is that of the B97M-V and B97M-rV functionals.
These functionals were the best performing semi-local density functionals in the H2Bind275 dataset (ranked 5\textsuperscript{th} and 6\textsuperscript{th}), and were recommended as inexpensive alternatives to the best performing and expensive density functionals.\cite{Veccham2020}
However, their performance in the current H2Bind78$\times$7 dataset deteriorates with B97M-rV and B97M-V yielding errors of 9.0\% and 9.1\% (ranked 14\textsuperscript{th} and 15\textsuperscript{th}) respectively.
While this reflects their lacklustre performance for the non-equilibrium geometries, they still remain the best performing semi-local functionals.
The next best performing semi-local density functional, revTPSS, is ranked 21\textsuperscript{st} and shows a RegMAPE of 12.0\%.
The performance of PBE0 and B2PLYP-D3(BJ) DFAs shows significant improvement relative to their performance in the H2Bind275 dataset with both density functionals entering the top 10 category for the H2Bind78$\times$7 dataset.

\begin{figure}
    \centering
    \includegraphics[width=\textwidth]{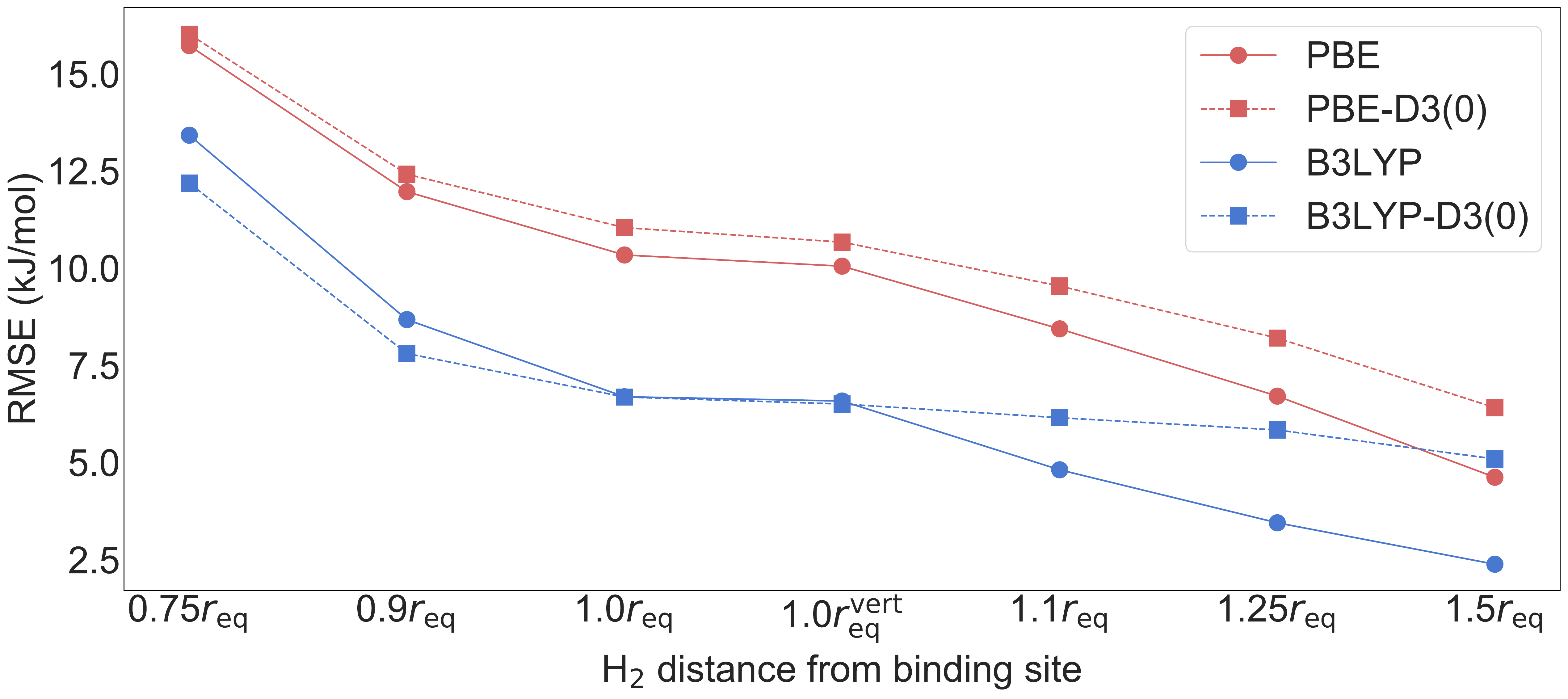}
    \caption{Effect of addition of empirical dispersion corrections on the RMSE of overbinding (PBE) and underbinding (B3LYP) density functionals at different points on the potential energy curve.}
    \label{fig:dispersion_effect}
\end{figure}
Other general trends also hold for DFAs assessed with the RegMAPE error metric.
The best DFA of each rung of the Jacob's ladder outperforms the best DFA from the rung below it.
The best performing meta-GGA functional is B97M-rV with a RegMAPE of 9.0\% and the best GGA is PBE with a RegMAPE of 15.3\%.
The effect of addition of empirical dispersion correction can also be assessed using the mean signed error (MSE) and RegMAPE metrics.
Addition of dispersion correction improves the performance only if the parent density functional has a systematic underbinding problem (characterized by a positive value of MSE).
For example, B3LYP has an MSE of 2.7 kJ/mol and a RegMAPE of 15.5\% and is systematically underbinding \ce{H2}(s).
Addition of a dispersion correction to B3LYP leads to the B3LYP-D3(0) functional which overcomes this underbinding problem.
B3LYP-D3(0) slightly overbinds with an MSE of $-0.8$ kJ/mol, but shows an improved RegMAPE of 14.0\%.
Addition of DFT-D corrections also improves the performance of other underbinding DFAs like revPBE and BLYP. 
However, addition of these corrections to parent DFAs that are already overbinding exacerbates the overbinding issue leading to poorer performance as exemplified by PBE, PBE0, TPSS, SCAN, and MS2 functionals.
Remarkably, PBE without any dispersion correction is the best performing GGA.
As empirical dispersion correction is distance dependent, it is interesting to see its effect at different points on the PEC.
For overbinding functionals, addition of dispersion correction worsens their performance across the PEC as exemplified by the PEC of PBE and PBE-D3(0) in Fig.~\ref{fig:dispersion_effect}.
The difference between the RMSEs of PBE and PBE-D3(0) increases with increasing distance of \ce{H2} with the binding site as dispersion corrections are usually damped in the short range.
Dispersion corrections improve the performance of underbinding functionals in the short range.
However, in the long range, dispersion corrections overestimate its magnitude, causing BLYP, B3LYP, and revPBE to overbind in the elongated regime.
MSEs and RMSEs of all the DFAs containing dispersion corrections and their corresponding parent functionals is shown in Table~S3.

\begin{figure}
    \centering
    \includegraphics[width=\textwidth]{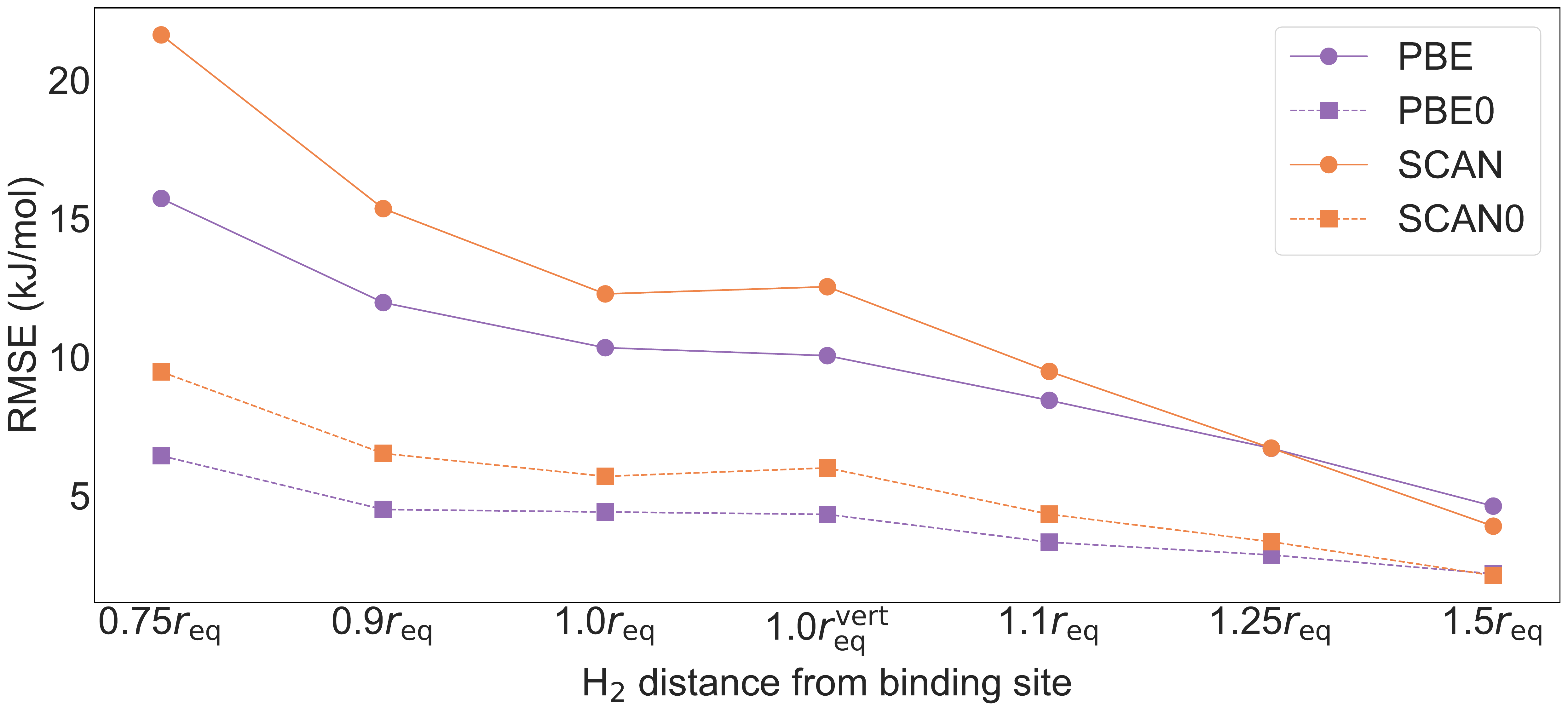}
    \caption{Performance of density functionals of the same family with and without Hartree Fock exchange at different points on the potential energy curve.}
    \label{fig:exact_exchange_effect}
\end{figure}
Addition of HF exact exchange is essential to ameliorate the effect of self interaction error in density functionals.
Comparing DFAs belonging to the same family, addition of HF exchange improves the performance of semi-local functionals for the H2Bind78$\times$7 dataset.
PBE0, which contains 25\% HF exchange, is ranked the 7\textsuperscript{th} with a RegMAPE of 7.6\%.
In contrast, the PBE functional is ranked 29\textsuperscript{th} with a RegMAPE for 15.3\%, more than two times that of PBE0.
HF exchange is a short-range effect and addition of HF exchange improves the performance of density functionals in the short range as shown in Fig.~\ref{fig:exact_exchange_effect}.
While the hybrid functional performs better than its semi-local counterpart throughout the PEC, its effect is more pronounced in the compressed region than the elongated region.
The SCAN and SCAN0 functionals show RMSEs of 21.6 and 9.5 kJ/mol (a difference of 12.2 kJ/mol) at $0.75r_{\text{eq}}$ of the PEC.
Their RMSEs at $1.5r_{\text{eq}}$ is 3.9 and 2.1 kJ/mol, with the hybrid functional improving on the semi-local one by only 1.8 kJ/mol.

The RegMAPE error metric gives larger weights to data points whose reference interaction energies are in the interesting range for \ce{H2} storage.
Further, it gives more weight to the equilibrium data point than non-equilibrium data points.
The relative weights of data points on the PES can be further tuned in order to assess the origin of errors of different DFAs.
Elongated geometries are encountered more often than compressed geometries in porous material capable of storing \ce{H2}.
Compressed geometries are also much higher in energy (geometries compressed by 25\% are almost always repulsive) and are encountered less often in simulations.
This would suggest retuning the weights of the regularized errors by giving larger weights to equilibrium and elongated regions of the PEC.
The weighted RegMAPE (denoted as wRegMAPE or $ \Delta E_w$) is defined as:
\begin{align}
     \Delta E_w &= \sum_i \frac{w_i \Delta  E(\alpha_i r_{\text{eq}})}{7}  \quad \text{s.t.} \sum_i w_i = 7  \label{eq:weighted_regmape}
\end{align}
where $ \Delta E(\alpha_i r_{\text{eq}})$ is the RegMAPE at the point $\alpha_i r_{\text{eq}}$ defined in Eq.~\eqref{eq:regMUPE_noneq}.
Ensuring that the weights sum up to $7$ would enable an apples-to-apples comparison of wRegMAPE and RegMAPE.
In the case of RegMAPE, $w_i=1$, for all values of $i$.
Reducing the weights of the compressed geometries with the scheme shown in Table~\ref{tab:weights}, the wRegMAPE can be computed using Eq.~\eqref{eq:weighted_regmape}.
This wRegMAPE metric, shown in Table~\ref{tab:wregmape}, gives more weight to the elongated geometries.
As the vertical interaction is computed its respective PEC minimum, the 1.0$r_{\text{eq}}^{\text{vert}}$ data point is assigned a weight equal to that of the adiabatic interaction energy at PEC minimum.

\begin{table}[]
\caption{Weights for different points on the adiabatic PEC and vertical interaction energy for calculating the weighted regularized mean absolute percentage error (wRegMAPE) metric.}
\label{tab:weights}
\begin{tabular}{|c|ccccccc|}
\hline
PEC location & $0.75r_{\text{eq}}$ & 0.9$r_{\text{eq}}$  & 1.0$r_{\text{eq}}$ & 1.0$r_{\text{eq}}^{\text{vert}}$    & 1.1$r_{\text{eq}}$  & 1.25$r_{\text{eq}}$ & 1.5$r_{\text{eq}}$  \\ \hline
Weight ($w_i$)       & 0.75 & 0.91 & 1.07 & 1.07 & 1.07 & 1.07 & 1.07 \\ \hline
\end{tabular}
\end{table}


\begin{table}[]
\caption{Weighted regularized mean absolute percentage error (wRegMAPE) for selected density functional approximations.}
\label{tab:wregmape}
\begin{tabular}{ccc}
\hline
Rank	&	DFA	&	wRegMAPE (\%)	\\ \hline
1	&	PBE0-DH	&	4.8	\\
2	&	$\omega$B97X-V	&	5.2	\\
3	&	DSD-PBEPBE-D3(BJ)	&	5.9	\\
4	&	$\omega$B97M-V	&	6.0	\\
5	&	XYGJ-OS	&	6.5	\\
6	&	$\omega$B97M(2)	&	7.0	\\
7	&	PBE0	&	7.4	\\
8	&	HSE-HJS	&	7.4	\\
9	&	XYG3	&	7.9	\\
10	&	B2PLYP-D3(BJ)	&	8.0	\\
11	&	B97M-rV	&	8.4	\\
12	&	B97M-V	&	8.4	\\
13	&	$\omega$B97X-D	&	8.5	\\
14	&	SCAN0	&	8.7	\\
15	&	PTPSS-D3(0)	&	9.1	\\ \hline
\end{tabular}%
\end{table}

Comparing the magnitude of the wRegMAPE (Table~\ref{tab:wregmape}) of different DFAs to their corresponding RegMAPE (Table~\ref{tab:err_regmape_rmse}), we can notice that the wRegMAPEs are slightly smaller.
Smaller wRegMAPEs suggest that density functionals perform better for the equilibrium and elongated geometries in comparison to the compressed ones.
However, the relative ordering of density functionals remains more or less the same.
The top five best performing DFAs (PBE0-DH, $\omega$B97X-V, $\omega$B97M-V, DSD-PBEPBE-D3(BJ), and XYGJ-OS) according to the RegMAPE metric are also the five best performing functionals according to the wRegMAPE metric.
We see that the recently parametrized $\omega$B97M(2) double hybrid density functional, which uses $\omega$B97M-V orbitals, is ranked sixth with wRegMAPE of 7.0\%.

\subsection{Performance of Density Functional Approximation for Geometries}
\begin{figure}
    \centering
    \includegraphics[width=\textwidth]{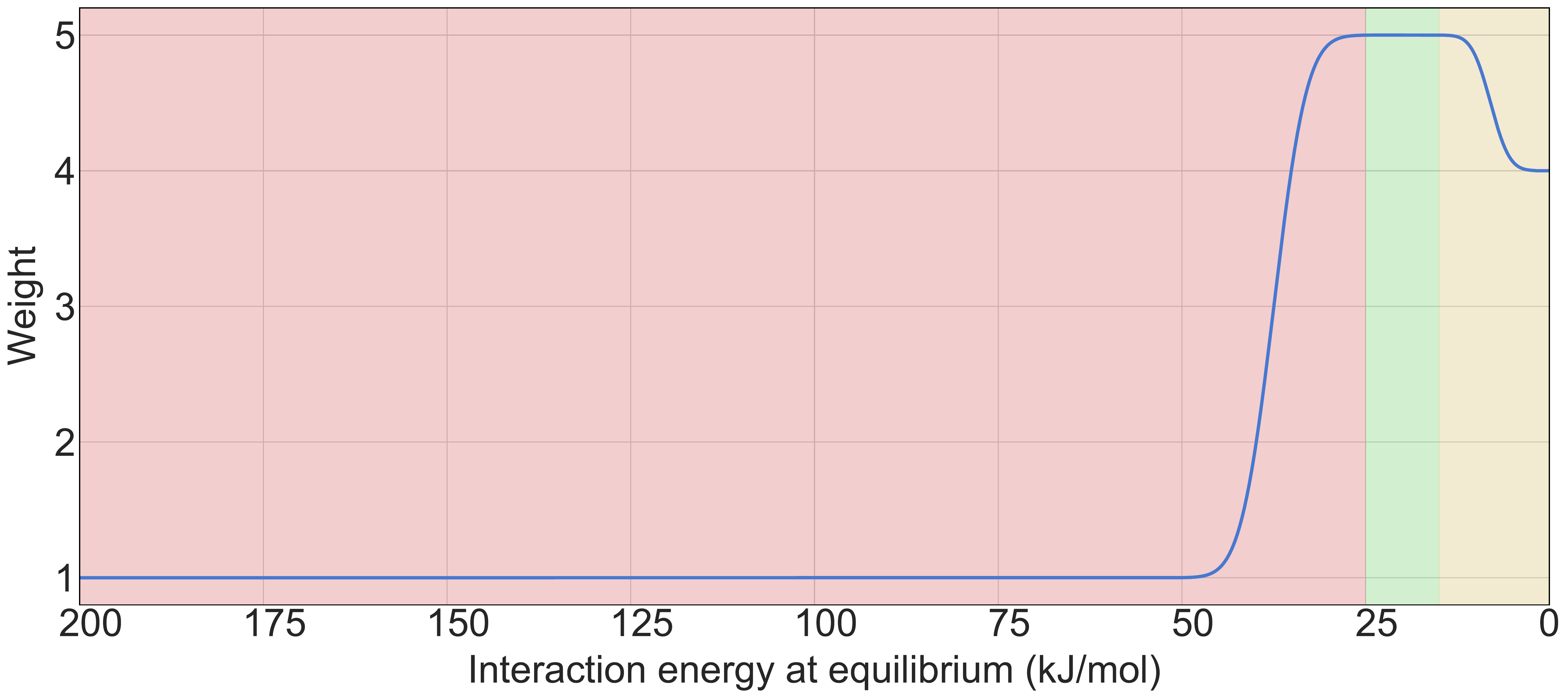}
    \caption{Weights of different chemical species as a function of their reference adiabatic interaction energy at equilibrium.}
    \label{fig:weights_assym}
\end{figure}

All the equilibrium geometries for this dataset were obtained by geometry optimization using the $\omega$B97M-V density functional in the def2-TZVPD basis set.\cite{Rappoport2010a}
With the exception of two chemical systems (\ce{AlF3-H2} and \ce{Ti+-(H2)2}), the CCSD(T)/CBS equilibrium geometry of all other chemical systems coincides (up to sampling precision) with the $\omega$B97M-V/def2-TZVPD equilibrium geometry.
This further validates the choice of equilibrium geometries for the H2Bind78$\times$7 dataset.

In order to assess the error in prediction of equilibrium geometry in a manner that is sensitive for \ce{H2} storage purposes, we have devised a weighting scheme that gives larger weights to more relevant data points.
Data points with reference adiabatic interaction energy at equilibrium ($E^{\text{ref}}(1.0 r_{\text{eq}})$) in the range of $-15$ to $-25$ kJ/mol are given a weight of $5.0$.
Equilibrium interaction energies stronger than $-25$ kJ/mol are assigned a weight of $1.0$.
These weights were chosen to reflect the relative importance of these data points in the RegMAPE metric.
In the RegMAPE metric, a density functional yielding an error of 1 kJ/mol in the strong binding regime contributes 1 unit to the total error as absolute error metric is used in this regime.
A DFA with an error of the 1 kJ/mol in the middle of the favorable regime for \ce{H2} storage (that is at $-20$ kJ/mol) contributes 5 units to the total error as percentage error metric is used.
The RegMAPE metric assigns a weight that is 5 times larger to the species in the favorable regime in comparison to the strong binders, thus justifying the weights of $5.0$ and $1.0$ in Fig.~\eqref{fig:weights_assym}.
The weak binders with equilibrium interaction energies weaker than $-15$ kJ/mol are mostly comprised of the organic ligands.
These species are ubiquitously found in porous materials capable of adsorbing \ce{H2}   (like MOFs) and form secondary binding sites for \ce{H2}.
As this regime is not as important as the favorable one, it is assigned a weight of $4.0$.
This weighting scheme is used to form the weighted mean signed error (wMSE) and weighted mean unsigned error (wMAE) metrics in Fig.~\eqref{fig:weights_assym}.
The reference adiabatic interaction energy at equilibrium decides the weight of the corresponding PEC.
As the vertical interaction energy does not lie on the adiabatic PEC, those data points were not included in the analyses in this section.

\begin{figure}
    \centering
    \includegraphics[width=\linewidth]{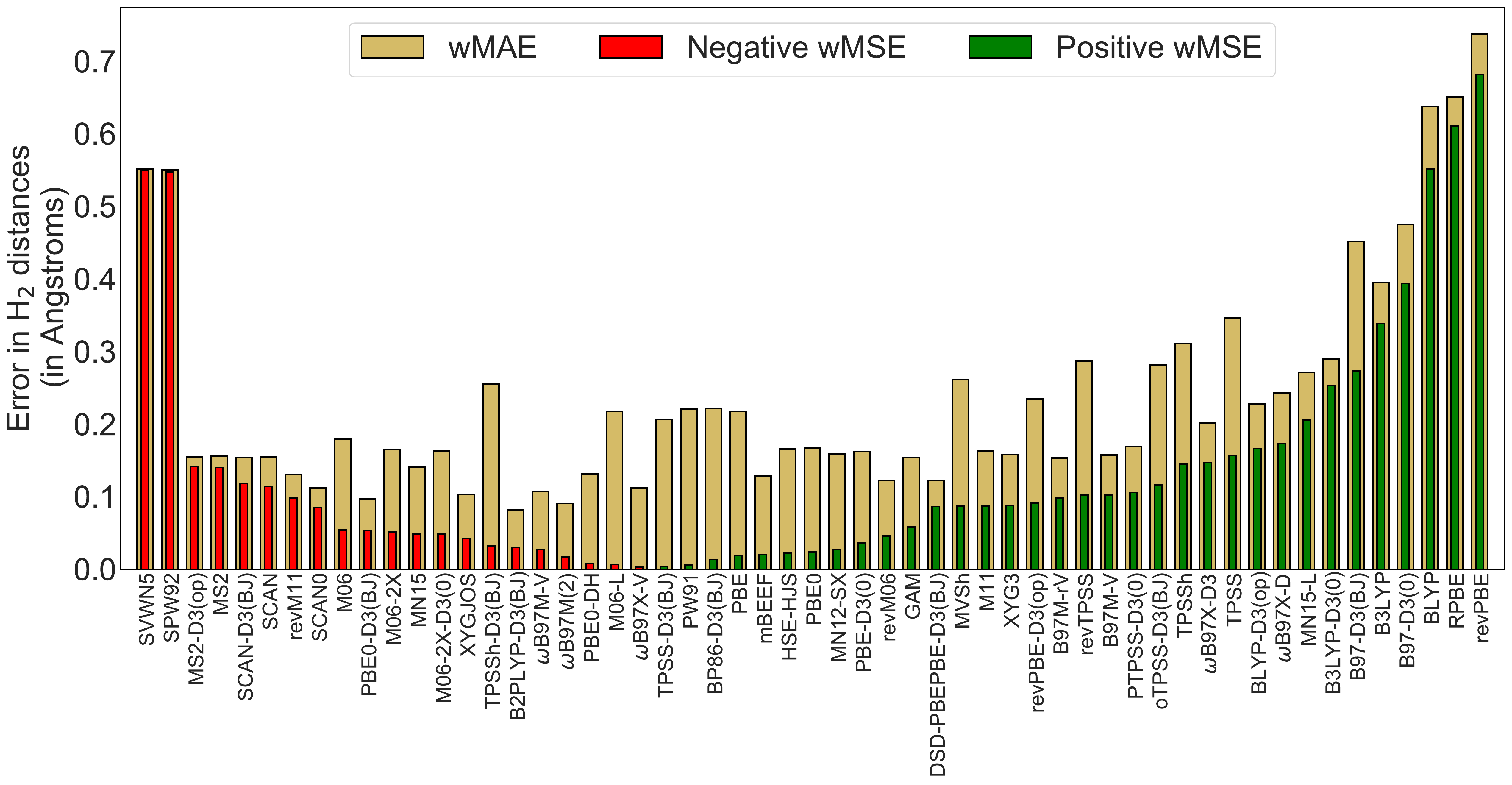}
    \caption{Weighted mean absolute error (wMAE) and weighted mean signed error (wMSE) of equilibrium \ce{H2} distances predicted by different DFAs.}
    \label{fig:H_displacements}
\end{figure}

Most DFAs predict longer equilibrium binding motif -- \ce{H2} distances which is shown as a positive value of wMSE in Fig.~\ref{fig:H_displacements}.
With the exception of the LDA density functionals, we see that all other density functionals which predict shorter equilibrium \ce{H2} distances have a very small negative wMSE ($> -0.15$\AA).
The double hybrid density functional B2PLYP-D3(BJ) gives the best performance for predicting equilibrium geometry with a wMAE of 0.08\AA.
It is closely followed by the recently parametrized double hybrid $\omega$B97M(2) density functional with a wMAE of 0.09\AA.
Both of these DFAs perform much better for equilibrium geometries in comparison to their performance for PECs.
It is also rather surprising to note the less good performance of PBE0-DH (wMAE of $0.13$\AA), which is the best performing DFA for binding energies according to RMSE and RegMAPE.
On the other hand, PBE0-D3(BJ) gives good geometries (ranked 3\textsuperscript{rd} with wMAE of 0.10 \AA) but its performance for PECs is mediocre (ranked 19\textsuperscript{th} with a RegMAPE of 11.5\%).
However, other top performing DFAs in the energetics category like XYGJ-OS, $\omega$B97M-V, and $\omega$B97X-V also perform well for geometries giving low wMAEs of 0.10 \AA, 0.11 \AA, and 0.11\AA~ respectively.
In particular, $\omega$B97X-V shows no systematic error with virtually zero wMSE.
It is also interesting to note the good performance of some semi-local density functionals like mBEEF and B97M-rV which give very low errors despite having no HF exchange.
In light of these observations, and given the enhanced computational cost of double hybrid DFA nuclear gradients,\cite{Neese2007analytic} one can use hybrid DFAs like PBE0-D3(BJ) or $\omega$B97M-V to perform a geometry optimization and then use the optimized geometry to perform a single point interaction energy calculation using a hybrid or double hybrid functional.

Another noticeable tread is the performance of DFAs upon the addition of some form of empirical dispersion correction.
Addition of empirical dispersion corrections to DFAs reduces their errors for equilibrium \ce{H2} distance prediction when the parent functional overestimates it.
For example, the addition of the D3(op) correction to revPBE decreases its wMSE from 0.68\AA~ to 0.09\AA~ (concurrently decreasing wMAE from 0.74\AA~to 0.23\AA).
The performance of the commonly used density functional B3LYP and M06-2X is quite poor with large wMSE and wMAEs.

\begin{table}[]
\caption{Performance of density functional approximations (DFAs) for predicting \ce{H2} binding energy at equilibrium geometry. The adiabatic regularized mean absolute percentage error (RegMAPE\textsubscript{ad}) for 15 best performing DFAs and some commonly used DFAs are shown.}
\label{tab:dft_mixed_regmupe}
\begin{tabular}{ccc}
\hline
Rank	&	DFA	&	RegMAPE\textsubscript{ad}	\\ \hline
1	&	$\omega$B97X-V	&	4.7	\\
2	&	DSD-PBEPBE-D3(BJ)	&	4.7	\\
3	&	PBE0-DH	&	5.3	\\
4	&	$\omega$B97M-V	&	6.0	\\
5	&	XYGJ-OS	&	7.1	\\
6	&	B97M-rV	&	7.2	\\
7	&	B97M-V	&	7.2	\\
8	&	$\omega$B97M(2)	&	7.4	\\
9	&	$\omega$B97X-D	&	7.8	\\
10	&	XYG3	&	7.9	\\
11	&	B2PLYP-D3(BJ)	&	8.3	\\
12	&	PBE0	&	8.4	\\
13	&	HSE-HJS	&	8.4	\\
14	&	$\omega$B97X-D3	&	9.5	\\
15	&	SCAN0	&	9.7	\\
28	&	B3LYP	&	15.7	\\
30	&	PBE	&	16.3	\\
31	&	SCAN	&	16.4	\\
33	&	revPBE-D3(op)	&	17.3	\\
42	&	mBEEF	&	19.9	\\
48	&	M06-2X	&	22.3	\\
51	&	B97-D3(BJ)	&	24.4	\\
52	&	GAM	&	26.5	\\
53	&	B97-D3(0)	&	29.4	\\ \hline
\end{tabular}
\end{table}

Typically, DFAs are used to optimize geometries of complexes containing an \ce{H2} bound to a binding motif.
After the geometry optimization has converged to a minimum on the potential energy surface, the binding energy of \ce{H2} is computed as the difference between the energy of the complex at the minimum of the potential energy surface and energies of the binding motif and \ce{H2} in isolation.
Alternatively, DFAs can also be used in molecular dynamics and Monte Carlo simulations either directly\cite{fetisov2018first} or indirectly (as reference energies for parametrizing force fields).\cite{Fang2013first,Fang2014, Becker2017, Dubbeldam2019design}
In these typical use cases, error in equilibrium binding energy can be attributed to two sources: (1) Inaccurate prediction of equilibrium geometry (2) Incorrect prediction of binding energy for the equilibrium geometry.
Using DFAs for modeling \ce{H2} binding materials, the DFA equilibrium geometry is typically used for computing the equilibrium binding energy.

We can assess the effect of relaxing the geometry along the PEC (defined by each DFA), the interaction coordinate of \ce{H2} with the binding site.
Geometry optimization along this coordinate can either improve or deteriorate the performance of density functionals.
The performance of selected density functionals for predicting the minimum energy geometry on its PEC and the \ce{H2} binding energy for this geometry assessed by the RegMAPE error metric is shown in Table~\ref{tab:dft_mixed_regmupe} (the complete table showing the performance of all 55 DFAs is included in Table~S4).

As both the reference and DFA geometries are relaxed along the potential energy curve, this error metric is adiabatic (ad) in nature as reflected in its subscript (RegMAPE\textsubscript{ad}).
The top five density functionals by the RegMAPE\textsubscript{ad} error metric are also the top five best performers according to their RegMAPE errors (Table~\ref{tab:err_regmape_rmse}), further emphasizing the superior performance of these DFAs for computing \ce{H2} interaction energies.
While the top five density functionals remain the same, it is interesting to note small changes in their order.
$\omega$B97X-V is the best performing DFA with a RegMAPE\textsubscript{ad} of 4.65\% which is very closely followed by DSD-PBEPBE-D3(BJ) with a RegMAPE\textsubscript{ad} of 4.68\%.
Another noteworthy difference is the performance of the B97M-rV and the B97M-V functionals which are ranked sixth and seventh with RegMAPE\textsubscript{ad} of 7.22\% and 7.23\% respectively.
These DFAs were ranked 12\textsuperscript{th} and 13\textsuperscript{th} with RegMAPE of 9.0\% and 9.1\%.
These functionals show a favorable cancellation of error in prediction of \ce{H2} equilibrium binding energies when the equilibrium geometry is also optimized using the same functional.
As these density functionals also do not have any HF exchange, they are computationally less expensive making them well-suited for applications in high-throughput material screening.
The rVV10 non-local functional, which is an approximation\cite{Sabatini2013} of the VV10 non-local functional, also allows for efficient evaluation in a plane wave framework and can be useful for modeling periodic systems like MOFs.
A thorough assessment of the DFA geometry relaxed on the entire potential energy surface, not just along the one-dimensional PEC, is beyond the scope of this work and we refer interested readers to Ref.~\citenum{Witte2015} for a detailed discussion of this topic.

\section{Conclusions}
The H2Bind275 dataset published recently\cite{Veccham2020} assesses the performance of density functionals for predicting the interaction energy of \ce{H2} with different model binding motifs at the  equilibrium geometry.
In this work, we have assessed the ability of DFAs to predict \ce{H2} interaction energies with binding motifs accurately throughout the PEC, not just at equilibrium geometry.
To that end, we have extended our previous H2Bind275 dataset by adding two compressed and three elongated geometries along the PEC to form the H2Bind78$\times$7 dataset.
The H2Bind78$\times$7 dataset comprises 545 data points at different fixed points along 78  PECs of various model binding motifs with \ce{H2}.
Reference interaction energies for all data points were computed using CCSD(T) extrapolated to the complete basis set limit.
The performance of 55 DFAs was assessed with the CCSD(T) reference interaction energies using multiple error metrics.
The RMSE metric is democratic and gives equal important to all the 545 data points.
The RegMAPE metric, on the other hand, gives more weight to binding motifs with interaction energies in the range of $-15$ to $-25$ kJ/mol at equilibrium geometry.
For each binding motif, the RegMAPE metric is designed to give more weight to the equilibrium than non-equilibrium data points as the latter are encountered less often in modeling and simulation.
DFAs are also assessed on the basis of their predicted equilibrium geometry and binding energy at predicted equilibrium geometry.

The CCSD(T) reference interaction energies for the H2Bind78$\times$7 dataset span a wide range of attractive and repulsive interaction energies.
As repulsive geometries are usually not included in non-covalent interaction energy datasets, the H2Bind78$\times$7 dataset adds considerably to the diversity of the available datasets.
The non-empirical double hybrid functional, PBE0-DH, shows the least error (RMSE of 2.9 kJ/mol and RegMAPE of 5.0\%) in predicting \ce{H2} binding energy throughout the PEC.
The $\omega$B97X-V, $\omega$B97M-V, and DSD-PBEPBE-D3(BJ) density functionals are also top performers.
The semi-local density functionals, B97M-V and B97M-rV, show poorer performance for the H2Bind78$\times$7 dataset, in comparison to the previous H2Bind275 dataset using the RegMAPE error metric.
For the H2Bind78$\times$7 dataset, B97M-V and B97M-rV are ranked 13\textsuperscript{th} and 12\textsuperscript{th} respectively with RegMAPEs of 9.1\% and 9.0\%.
Previously in the H2Bind275 dataset, they were ranked 5\textsuperscript{th} and 6\textsuperscript{th} with RegMAPE of 6.8\%.
In general, the good performance of the top density functionals in the H2Bind275 dataset continues for the H2Bind78$\times$7 dataset.
Addition of DFT-D empirical dispersion correction increases the accuracy of underbinding density functionals like revPBE, BLYP, and B3LYP.
This addition also decreases the accuracy of overbinding parent density functionals like PBE, PBE0, TPSS, SCAN, and MS2.
As DFT-D empirical dispersion corrections are distance dependent, their effect is not felt uniformly across the PEC.
The effect of addition of HF exchange, a short-ranged effect, improves the performance of density functionals in the compressed regime more than in the elongated regime of the PEC, thus playing a crucial role in accurately predicting the repulsive wall of the PEC. 
The weighted RegMAPE metric gives smaller weights to DFA errors in the compressed region of the PEC. This metric shows that, in general, DFAs perform better in the equilibrium and elongated regime than in the compressed region.

Assessment of DFAs for predicting equilibrium geometries reveals that PBE0-DH, which is the best performer for energies, shows less good performance for geometries.
However, other hybrid functionals like $\omega$B97M-V and $\omega$B97X-V give good performance for both geometries and energies.
Using the adiabatic RegMAPE metric (RegMAPE\textsubscript{ad}) reveals that the semi-local DFAs, B97M-V and B97M-rV, show very small errors.
They benefit significantly from cancellation between geometry-driven and energy-driven errors.
$\omega$B97M-V and $\omega$B97X-V are the only density functionals that are not double hybrids which consistently show good performance for all the error metrics (energy and geometry-related) defined in this work.
As these hybrid functionals have significantly lower computational cost in comparison to double hybrids, we recommend their usage for \ce{H2} binding applications.

The H2Bind78$\times$7 dataset, consisting of highly accurate reference interaction energies, represents a distinctive addition to other non-covalent interaction energy databases.
More than half of this dataset consists of transition metal species which are usually underrepresented in non-covalent interaction energy datasets.
This dataset is composed of complete PECs, rather than just PEC minimum geometries -- only a handful of the non-covalent interaction energy datasets contain this information.
Besides, almost all of the PECs in this dataset sample the repulsive wall.
For these reasons, using the H2Bind78$\times$7 dataset in training or validating DFAs can improve their performance and transferability.
This work further validates the selection of best performing density functionals for the H2Bind275 dataset using a semi-independent dataset that is about two times larger.
The definition and generalization of different error metrics (RegMAPE, wRegMAPE, and RegMAPE\textsubscript{ad}) can be used for assessment of other similar datasets with well-defined schemes for weighting different data points.
As force field parametrization requires good reference energies throughout the potential energy surface, the top performing density functionals in this work can be used for generating them.

\begin{acknowledgement}
This work was supported by the Hydrogen Materials - Advanced Research Consortium (HyMARC), established as part of the Energy Materials Network under the U.S. Department of Energy, Office of Energy Efficiency and Renewable Energy, under Contract No. DE-AC02-05CH11231. The following author declares a competing financial interest. M. H. G. is a part owner of Q-Chem, Inc.
\end{acknowledgement}

\begin{suppinfo}
Additional information regarding the parameters used for double hybrid density functionals, performance of density functionals for the non-equilibrium subset of the H2Bind78$\times$7 dataset, and the RegMAPE\textsubscript{ad} error for all DFAs assessed in this work is included in the supplementary information.
CCSD(T)/CBS and DFA interaction energies for all 55 functionals are provided in the file supporting\_information.xlsx.
The geometries of all complexes in the H2Bind78$\times$7 are included in geometries.zip file.
\end{suppinfo}

\bibliography{references}
\end{document}


\begin{table}[H]
\caption{Parameters for treatment of MP2 correlation energy for all double hybrid functionals considered in this work.}
\begin{tabular}{ccc}
\hline
Density functional & Density fitting employed? & Level of frozen core approximation    \\ \hline
$\omega$B97M(2)    & yes                       & as recommended by Ref.~\citenum{Mardirossian2018}                    \\
DSD-PBEPBE-D3(BJ)  & yes                       & all core electrons                    \\
XYG3               & no                        & all core electrons                    \\
B2PLYP-D3(BJ)      & yes                       & no core electrons were frozen         \\
PTPSS-D3(0)        & yes                       & all core electrons                    \\
PBE0-DH            & no                        & no core electrons were frozen         \\
XYGJOS             & yes                       & same as the coupled cluster reference \\ \hline
\end{tabular}
\end{table}

\begin{table}[H]
\caption{Performance of all 55 density functional approximations (DFAs) for non-equilibrium geometries ($0.75r_{\text{eq}}$, $0.9r_{\text{eq}}$, $1.1r_{\text{eq}}$, $1.25r_{\text{eq}}$, and $1.5r_{\text{eq}}$)}
\label{tab:reg_err_x}
\resizebox{0.8\textwidth}{!}{%
\begin{tabular}{|c|cc|cc|}
\hline
Rank	&	DFA	&	RegMAPE (\%)	&	DFA	&	RMSE (in kJ/mol)	\\ \hline
1	&	PBE0-DH	&	4.7	&	PBE0-DH	&	2.8	\\
2	&	$\omega$B97X-V	&	5.6	&	DSD-PBEPBE-D3(BJ)	&	3.9	\\
3	&	$\omega$B97M-V	&	6.2	&	PBE0	&	4.1	\\
4	&	XYGJ-OS	&	6.7	&	$\omega$B97X-V	&	4.2	\\
5	&	DSD-PBEPBE-D3(BJ)	&	7.0	&	$\omega$B97X-D	&	4.2	\\
6	&	PBE0	&	7.0	&	HSE-HJS	&	4.2	\\
7	&	HSE-HJS	&	7.1	&	MVSh	&	4.4	\\
8	&	$\omega$B97M(2)	&	7.2	&	XYGJ-OS	&	4.6	\\
9	&	B2PLYP-D3(BJ)	&	8.0	&	$\omega$B97M-V	&	4.7	\\
10	&	XYG3	&	8.6	&	PTPSS-D3(0)	&	4.9	\\
11	&	SCAN0	&	8.7	&	XYG3	&	4.9	\\
12	&	PTPSS-D3(0)	&	9.1	&	$\omega$B97X-D3	&	5.0	\\
13	&	$\omega$B97X-D	&	9.4	&	$\omega$B97M(2)	&	5.1	\\
14	&	B97M-rV	&	9.6	&	PBE0-D3(BJ)	&	5.1	\\
15	&	B97M-V	&	9.7	&	MN15	&	5.5	\\
16	&	$\omega$B97X-D3	&	9.9	&	SCAN0	&	5.8	\\
17	&	MVSh	&	10.4	&	B2PLYP-D3(BJ)	&	5.8	\\
18	&	TPSSh	&	10.5	&	TPSSh	&	6.2	\\
19	&	revTPSS	&	11.0	&	revM11	&	6.5	\\
20	&	PBE0-D3(BJ)	&	11.3	&	mBEEF	&	6.8	\\
21	&	revM06	&	12.0	&	revM06	&	7.1	\\
22	&	M11	&	12.5	&	revTPSS	&	7.6	\\
23	&	TPSS	&	12.6	&	B3LYP	&	7.7	\\
24	&	B3LYP-D3(0)	&	14.0	&	TPSS	&	7.8	\\
25	&	MN15	&	14.4	&	B97M-rV	&	7.8	\\
26	&	TPSSh-D3(BJ)	&	14.4	&	B97M-V	&	7.8	\\
27	&	TPSS-D3(BJ)	&	14.4	&	B3LYP-D3(0)	&	7.9	\\
28	&	oTPSS-D3(BJ)	&	14.4	&	oTPSS-D3(BJ)	&	7.9	\\
29	&	revM11	&	14.4	&	MN15-L	&	8.3	\\
30	&	PBE	&	14.6	&	TPSS-D3(BJ)	&	8.5	\\
31	&	BLYP-D3(op)	&	15.1	&	TPSSh-D3(BJ)	&	8.5	\\
32	&	B3LYP	&	15.2	&	revPBE-D3(op)	&	8.7	\\
33	&	SCAN	&	15.3	&	MN12-SX	&	8.9	\\
34	&	MN12-SX	&	15.9	&	revPBE	&	9.4	\\
35	&	SCAN-D3(BJ)	&	16.1	&	M11	&	9.5	\\
36	&	mBEEF	&	16.2	&	RPBE	&	9.6	\\
37	&	M06	&	16.3	&	B97-D3(BJ)	&	9.8	\\
38	&	revPBE-D3(op)	&	16.8	&	BLYP-D3(op)	&	10.1	\\
39	&	MS2	&	16.9	&	PBE	&	10.3	\\
40	&	PW91	&	17.0	&	BLYP	&	10.4	\\
41	&	BP86-D3(BJ)	&	17.8	&	M06-L	&	10.5	\\
42	&	M06-2X	&	18.4	&	M06	&	10.5	\\
43	&	M06-2X-D3(0)	&	18.9	&	BP86-D3(BJ)	&	10.9	\\
44	&	M06-L	&	19.7	&	B97-D3(0)	&	11.0	\\
45	&	MS2-D3(op)	&	19.8	&	PBE-D3(0)	&	11.1	\\
46	&	PBE-D3(0)	&	20.0	&	PW91	&	11.2	\\
47	&	MN15-L	&	20.1	&	GAM	&	11.6	\\
48	&	RPBE	&	20.8	&	MS2	&	12.0	\\
49	&	BLYP	&	21.6	&	MS2-D3(op)	&	12.2	\\
50	&	revPBE	&	22.9	&	SCAN	&	13.1	\\
51	&	GAM	&	23.9	&	SCAN-D3(BJ)	&	13.3	\\
52	&	B97-D3(BJ)	&	24.2	&	M06-2X	&	14.0	\\
53	&	B97-D3(0)	&	27.9	&	M06-2X-D3(0)	&	14.0	\\
54	&	SPW92	&	62.9	&	SPW92	&	32.8	\\
55	&	SVWN5	&	63.0	&	SVWN5	&	32.9	\\ \hline
\end{tabular}%
}
\end{table}

\begin{table}[H]
\caption{Performance of density functional approximations (DFAs) with and without empirical dispersion corrections shown with RegMAPE (\%), RMSE (in kJ/mol), MSE (in kJ/mol), and their ranking by RegMAPE metric among the 55 DFAs assessed.}
\label{tab:empirical_dispersion}
\resizebox{\textwidth}{!}{%
\begin{tabular}{ccccc|ccccc}
\hline
\multicolumn{5}{c|}{Parent DFA}        & \multicolumn{5}{c}{DFA with dispersion}            \\ \hline
DFA	&	RegMAPE	&	RMSE	&	MSE	&	Rank	&	DFA	with	dispersion	&	RegMAPE	&	RMSE	&	MSE	&	Rank	\\ \hline
PBE	&	15.3	&	10.3	&	-5.1	&	29	&	PBE-D3(0)	&	20.1	&	11.0	&	-7.1	&	45	\\
revPBE	&	23.4	&	9.1	&	3.8	&	50	&	revPBE-D3(op)	&	17.0	&	8.5	&	-3.5	&	37	\\
BLYP	&	22.1	&	10.1	&	2.4	&	49	&	BLYP-D3(op)	&	15.4	&	9.8	&	-2.0	&	31	\\
TPSS	&	13.6	&	7.7	&	-1.0	&	23	&	TPSS-D3(BJ)	&	14.7	&	8.5	&	-4.2	&	27	\\
SCAN	&	15.6	&	12.9	&	-6.7	&	33	&	SCAN-D3(BJ)	&	16.3	&	13.2	&	-7.3	&	35	\\
MS2	&	17.4	&	12.1	&	-7.1	&	39	&	MS2-D3(op)	&	20.3	&	12.4	&	-7.8	&	46	\\
B3LYP	&	15.5	&	7.4	&	2.7	&	32	&	B3LYP-D3(0)	&	14.0	&	7.5	&	-0.8	&	25	\\
PBE0	&	7.6	&	4.2	&	-1.3	&	7	&	PBE0-D3(BJ)	&	11.5	&	5.2	&	-3.5	&	19	\\
TPSSh	&	11.3	&	6.0	&	0.1	&	18	&	TPSSh-D3(BJ)	&	13.8	&	8.0	&	-3.7	&	24	\\
M06-2X	&	19.3	&	13.4	&	0.6	&	42	&	M06-2X-D3(0)	&	19.8	&	13.4	&	0.5	&	43	\\ \hline
\end{tabular}%
}
\end{table}

\begin{table}[H]
\caption{Performance of all 55 density functional approximations (DFAs) for predicting \ce{H2} binding energy at equilibrium geometry shown using the adiabatic regularized mean absolute percentage error (RegMAPE\textsubscript{ad}). }
\label{tab:regmape_ad_extrap_full}
\resizebox{0.42\textwidth}{!}{%
\begin{tabular}{|c|cc|}
\hline
Rank	&	DFA	&	RegMAPE\textsubscript{ad} (\%)	\\ \hline
1	&	$\omega$B97X-V	&	4.7	\\
2	&	DSD-PBEPBE-D3(BJ)	&	4.7	\\
3	&	PBE0-DH	&	5.3	\\
4	&	$\omega$B97M-V	&	6.0	\\
5	&	XYGJ-OS	&	7.1	\\
6	&	B97M-rV	&	7.2	\\
7	&	B97M-V	&	7.2	\\
8	&	$\omega$B97M(2)	&	7.4	\\
9	&	$\omega$B97X-D	&	7.8	\\
10	&	XYG3	&	7.9	\\
11	&	B2PLYP-D3(BJ)	&	8.3	\\
12	&	PBE0	&	8.4	\\
13	&	HSE-HJS	&	8.4	\\
14	&	$\omega$B97X-D3	&	9.5	\\
15	&	SCAN0	&	9.7	\\
16	&	MVSh	&	9.9	\\
17	&	PTPSS-D3(0)	&	10.4	\\
18	&	M11	&	11.1	\\
19	&	PBE0-D3(BJ)	&	11.7	\\
20	&	TPSSh-D3(BJ)	&	12.0	\\
21	&	TPSSh	&	12.3	\\
22	&	revTPSS	&	13.5	\\
23	&	revM06	&	13.7	\\
24	&	oTPSS-D3(BJ)	&	14.4	\\
25	&	TPSS	&	14.6	\\
26	&	TPSS-D3(BJ)	&	14.8	\\
27	&	BLYP-D3(op)	&	15.6	\\
28	&	B3LYP	&	15.7	\\
29	&	B3LYP-D3(0)	&	16.1	\\
30	&	PBE	&	16.3	\\
31	&	SCAN	&	16.4	\\
32	&	SCAN-D3(BJ)	&	17.2	\\
33	&	revPBE-D3(op)	&	17.3	\\
34	&	MN12-SX	&	17.3	\\
35	&	MN15	&	17.5	\\
36	&	PW91	&	17.9	\\
37	&	revM11	&	17.9	\\
38	&	RPBE	&	18.1	\\
39	&	MS2	&	19.1	\\
40	&	M06	&	19.3	\\
41	&	BP86-D3(BJ)	&	19.4	\\
42	&	mBEEF	&	19.9	\\
43	&	MN15-L	&	20.1	\\
44	&	PBE-D3(0)	&	20.6	\\
45	&	revPBE	&	21.6	\\
46	&	BLYP	&	21.6	\\
47	&	MS2-D3(op)	&	22.0	\\
48	&	M06-2X	&	22.3	\\
49	&	M06-2X-D3(0)	&	22.6	\\
50	&	M06-L	&	24.1	\\
51	&	B97-D3(BJ)	&	24.4	\\
52	&	GAM	&	26.5	\\
53	&	B97-D3(0)	&	29.4	\\
54	&	SPW92	&	73.9	\\
55	&	SVWN5	&	73.9	\\ \hline
\end{tabular}%
}
\end{table}

\bibliography{references}